# Observation of sequential quantum oscillations induced by mini-Landau bands in a three-dimensional Dirac semiconductor

Zezhi Wang[1,2*], Dong Xing[1,2*], Bingbing Tong[1], Senyang Pan[3], Guangtong Liu[1,4], Li Lu[1,4], Jinglei Zhang[3], and Cheng-Long Zhang[1†]

[1]*Beijing National Laboratory for Condensed Matter Physics, Institute of Physics, Chinese Academy of Sciences, Beijing 100190, China*

[2]*School of Physical Sciences, University of Chinese Academy of Sciences, Beijing 100049, China*

[3] *High Magnetic Field Laboratory, HFIPS, Chinese Academy of Sciences, Hefei 230031, China*

[4]*Hefei National Laboratory, Hefei 230088, China*

Corresponding author: chenglong.zhang@iphy.ac.cn

**Quantum oscillations, the oscillatory behavior of electrical and thermodynamic properties, are typically observed in metals and vanish in the quantum limit under strong magnetic fields[1]. Phenomena such as the fractional quantum Hall effect[2], the Hofstadter butterfly[3,4], and recent observations of quantum oscillations in exotic insulators are notable exceptions[5-12]. The narrow-gap Dirac semiconductor $ZrTe_5$, a less exotic material without strong correlations or artificially engineered superlattices, nevertheless exhibits resistance oscillations in the quantum limit[13] but can be interpreted within a simple Zeeman-effect-based picture[14,15], which remains conventional quantum oscillations without exotic properties. Here, we report the observation of unexpected mini-oscillations superimposed on Zeeman-effect-induced main oscillations in the quantum limit. The subtracted mini-oscillations are periodic in $1/B$ with the highest frequency equal to 2.1% of the first Brillouin zone and have extremely heavy effective mass ~ $2m_e$, which is unexpected in $ZrTe_5$ given its ultralow carrier density. Additionally, the mini-oscillations exhibit sequential features that are synchronized with the main oscillations, suggesting an internal structure of the Landau bands. However, they appear incompatible with the Hofstadter butterfly due to the highly anisotropic/three-dimensional crystal structure. These sequential mini-oscillations correlate with the commensurability resonance effect with subunity fractions observed in angular magnetoresistance, relating to the formation of mini-Landau bands. Our results present solid experimental evidence of exotic quantum oscillations in the quantum limit beyond currently available mechanisms, and establish $ZrTe_5$, a prototypical Dirac semiconductor, as a simple platform parallel to correlated insulators for exploring exotic oscillations.**

In an external magnetic field, a two-dimensional (2D) electron gas exhibits quantum oscillations (QOs) in density-of-states-related electrical and thermodynamic properties, due to Landau quantization. In other words, QOs should occur in metals with discrete Landau levels (LLs) on a well-defined Fermi surface[1]. QOs cease when the magnetic field approaches a critical value, known as the quantum limit, at which all LLs except the last one move across the Fermi level. However, some exotic QOs fall outside the paradigm described above. The fractional quantum Hall effect is a prominent example in which electron correlations lead to fractionalization of LLs, including the lowest LL[2]. Another notable example is the Hofstadter butterfly, which arises when a 2D electron gas is in a superlattice with a scale comparable to the magnetic length, leading to a fractal pattern of LLs[3,4]. Recent observations of QOs in topological/correlated insulators have sparked debates over different mechanisms, including modulations of the hybridized gap and the possible existence of exotic neutral fermionic excitations[5-12,16,17]. All the above-mentioned phenomena are observed in exotic systems, including nontrivial band insulators or metallic electron gas with correlations/artificial superlattices. The question is: could we observe exotic QOs in less exotic systems without the ingredients mentioned above?

$ZrTe_5$, as a typical weak topological insulator[18-20] (equal to a normal insulator by the classification of the main $Z_2$ index), is a narrow-gap semiconductor with a Dirac dispersion in an orthorhombic layered structure (a = 3.9875 Å, b = 14.530 Å, c = 13.724 Å)[21], which has been studied for over 50 years, initially for the anomaly of resistivity[22] and recently for its topological properties[23-26]. Flux-grown $ZrTe_5$ crystals exhibit ultralow carrier density ranges from $10^{14}$ to $10^{15}$ cm$^{-3}$, and ultrahigh mobility up to $10^6$ cm$^2$ V$^{-1}$s$^{-1}$ at 2 K[14,27-29]. Based on carrier density and anisotropy of the Fermi surface[24], the quantum limit is roughly around 0.05 ~ 0.25 T (the estimation is not precise if charge puddles are present[30]). Signatures of fractional peaks in resistance oscillations beyond the quantum limit have been reported in semimetals and three-dimensional (3D) topological insulators, but remain elusive and later proved to be caused by twin or sample-dependent[31-35]. Despite the absence of correlations, concrete experimental evidence for the appearance of QOs in the quantum limit has been observed in $ZrTe_5$[13,14]. The QOs observed in the quantum limit of $ZrTe_5$ can be explained by a less exotic physical picture, in which the Zeeman effect dominates over the square-root Dirac kinetic energy (field-enhanced effective mass), resulting in bending of the Landau bands (LBs) in the quantum limit[14,15]. This means the QOs observed in the quantum limit of $ZrTe_5$ remain conventional. However, the above-mentioned simple picture has a direct consequence: a field-induced insulator-metal transition is expected when LBs cross at high fields, which contradicts the experimental results (Fig. 1**a**&1**d**). This inconsistency suggests that exotic properties of QOs in $ZrTe_5$ remain unexplored.

Here, we report high-resolution magnetoresistance measurements in $ZrTe_5$ conducted at sub-Kelvin temperatures using high-field facilities. We unexpectedly uncover exotic QOs (named as mini-oscillations) parasitic on the previously discovered Zeeman-effect-induced oscillations (named as main oscillations) in the quantum limit. The mini-oscillations are periodic in $1/B$ with high frequencies, and sequentially synchronize with the main oscillations. The temperature-dependent amplitude of mini-oscillations yields an extremely heavy effective mass $\sim 2m_e$ around 30 T. Further experiments show that the mini-oscillations closely correlate with the commensurability resonance observed in angular magnetoresistance oscillations (AMROs), pointing to the formation of mini-LBs. Although the sequential behavior of the observed mini-oscillations indicates the existence of internal structures on LBs that resemble the Hofstadter butterfly in the 2D system, we should note that $ZrTe_5$ is a highly anisotropic 3D system.

As shown in Fig. 1**a**, measurements of magnetoresistance $\rho_{xx}$ were performed on a $ZrTe_5$ single crystal with carrier density $\sim 5 \times 10^{14}$ cm$^{-3}$ at 2 K, corresponding to a quantum limit $\sim 0.07$ T[14] (see Supplementary Fig. 1 for more details of structural characterizations and basic transport properties of sample W049), and the current-voltage and rotation configurations are defined in the inset of Fig. 1**a**. The previously reported Zeeman-effect-induced oscillations[14], named as main oscillations, can be clearly resolved in low magnetic fields (PPMS-9T) at $\theta = 70°$. The amplified region in Fig. 1**a** exhibits mini-oscillations in the raw data. In Fig. 1**b**, the background (see Supplementary Note 1 and Supplementary

Fig. 3 for details of background subtraction) was subtracted to show the main oscillations and superimposed mini-oscillations. In Fig. 1**c**, we further subtracted the slow main oscillations to obtain the signals of mini-oscillations. The mini-oscillations diminish rapidly with increasing temperature, showing a distinct temperature dependence from that of the main oscillations. To clearly show the mini-oscillations around θ = 0°, we performed measurements in a steady magnetic field of up to 35 T with a He-3 insert. As shown in Fig. 1**d**, the main oscillations are obvious, but the mini-oscillations are not visible in the raw data due to the strong magnetoresistance. This can be overcome by monitoring the second-harmonic response ($R_{2\omega}$), which is proportional to the thermoelectric signals arising from the small thermal gradient across the sample. As shown in Fig. 1**e**, we successfully observe the mini-oscillations in the raw data. In Figs. 1**f**-1**h**, we plot the subtracted main and mini-oscillations up to 35 T, and the mini-oscillations show faster decay at sub-Kelvin temperatures. Experimental exhibition of mini-oscillations in raw data, without subtraction, secures the main observation in this paper, which is also reproduced across different samples (W049 and S74 in the main text) and batches.

We now turn to the analysis of the peculiar temperature dependence of the mini-oscillations. Figures 2**a** and 2**b** show the temperature-dependent amplitudes of the main ($|\Delta\rho_{xx}^{\text{main}}|$) and mini-oscillations ($|\Delta\rho_{xx}^{\text{mini}}|$) measured up to 9 T, respectively. Figure 2**c** shows the temperature-dependent amplitudes of $|\Delta\rho_{xx}^{\text{mini}}|$, measured up to 29.6 T at sub-Kelvin temperatures. Data in Figs. 2**a**-2**c** are fitted by the temperature-damping prefactor $\lambda = \frac{2\pi^2 k_B m_c}{\hbar eB}T$ of Lifshitz–Kosevich (L–K) formula[1], where $k_B$ is the Boltzmann constant, $m_c$ is the effective mass, $e$ is the elementary electron charge, and $\hbar$ is the reduced Planck constant. Distinct L-K fittings are observed between the main and mini-oscillations, where the onset temperature ($T_{\text{onset}}^{\text{mini}}$) of the mini-oscillations is roughly estimated to be 25 K based on the L-K fitting. Figure 2**d** summarizes all the fitted effective masses for both the main and mini-oscillations. The effective mass of the mini-oscillations is one order heavier than that of the main oscillations, attaining values larger than the mass of free electrons ($m_e$) above 20 T, and then reveals a distinct origin compared to the main oscillation. Given that the normal effective mass (cyclotron mass in *ac* plane) of the Dirac electron in ZrTe$_5$ ~ $10^{-2}$-$10^{-3}$ $m_e$[24,36,37], the observed heavy effective mass (~ 2 $m_e$) indicates the mini-oscillation is exotic.

After distinguishing from the main oscillation, we now turn to know more detailed properties of these exotic mini-oscillations by inspecting their oscillatory patterns. In Fig. 3**a**, the main and mini-oscillations, subtracted from the angle-dependent magnetoresistance measured up to 9 T, show a traceable feature when the magnetic field rotates from the *b*- to *c*-axes. At θ ~ 60°, the mini-oscillations can be clearly resolved due to suppressed amplitudes of the main oscillations. Three obvious oscillatory sequences of mini-oscillations are marked by colored inverted triangles, squares and circles, respectively, meaning three QOs frequencies that occur sequentially without overlapping. We also note that the mini-oscillations are clearly resolved only in the *bc* plane (the absence of mini-oscillations rotated in the *ba* plane is shown in Supplementary Fig. 5**b**).

This special oscillatory pattern drives us to conduct the measurement in higher magnetic fields at sub-Kelvin temperatures. However, we encountered severe resolution issues in measuring such small wriggles (mini-oscillations) superimposed on the main oscillations with a 35 T water-cooled magnet. We then switched to a top-loading dilution refrigerator with an 18 T superconducting magnet and further enhanced the mini-oscillations by tilting the angle θ. As the main dataset in this paper, Figs. 3**b** and 3**c** show subtracted $\Delta\rho_{xx}$ versus $1/B$ at each angle θ, the mini-oscillations are superimposed on the main oscillations and clearly resolved, as indicated by markers. All the observed mini-oscillations are evenly distributed in $1/B$ (in contrast to the non-$1/B$-periodic main oscillations[14]), but exhibit two unconventional properties: *1*. the mini-oscillations exhibit four monotonically increasing frequencies; *2*. the mini-oscillations with different periodicities are locked to the main oscillations and appear in sequence. As shown in Fig. 3**d**, we assign arbitrary integers $N$ to the peaks of the sequential mini-oscillations; the resulting $N$ versus $1/B$ Landau fan diagram shows four sequentially appearing frequencies: $F_1 = 8$ T, $F_2 = 24.4$ T, $F_3 = 73.7$ T, and $F_4 = 161.1$ T. The interlocking behaviors between the main and mini-oscillations are clearly shown in the Landau fan diagram (Fig. 3**d**) when the main oscillations are included. As shown in Fig. 3**d**, the four colored regimes distinguish the four sequential mini-oscillations with distinct frequencies, we can find that the four colored regimes are almost coincident with one period (one peak and one valley) of the main oscillations. The two aforementioned unconventional properties indicate that mini-oscillations originate from the internal structures of the LBs (mini-Landau bands) and differ from the well-known commensurate modulation-induced band oscillations, namely Weiss oscillations[38,39]. The highest frequency $F_4$ equal to 2.1% occupation of the first Brillouin zone, while the ultralow bulk density ~ $10^{14}$ cm$^{-3}$ corresponds to $10^{-5}$ occupation of the first Brillouin zone. In Fig. 3**e**, four sequentially observed frequencies decrease when the angle θ increases, which is distinct from the anisotropy of the main oscillations[14]. The internal structures of the LBs seem to resemble the Hofstadter butterfly observed in 2D systems, with fine-tuned superlattice modulation[4]. However, the highly anisotropic 3D structure (large anisotropy in both in-plane and out-of-plane directions due to the quasi-one-dimensional (1D) ZrTe$_3$ chains along ***a***-axis) of ZrTe$_5$ makes it unlikely that the Hofstadter butterfly will appear here. We therefore conclude that the experimental observations represent a previously unknown QOs.

What is the origin of the internal structure of the LBs? Driven by the observation that the mini-oscillations are clearly resolved at a tilted angle θ within the ***bc*** plane, we suspect that they may be related to the tilted configuration. We then performed angular magnetoresistance (AMR) measurements on the same sample that shows sequential mini-oscillations. Figures 4**a** and 4**b** show the AMR measured in different magnetic fields, which exhibit typical features of AMROs (the lower panel of Fig. 4**a** shows the second derivatives). We find that the AMROs with field-independent dips are strongly correlated with the observed mini-oscillations; stronger mini-oscillations correspond to stronger AMROs dips, and samples without detectable AMROs also exhibit no mini-oscillations (see Supplementary Fig. 6). The AMROs are also absent in CVT-grown samples (see Supplementary Note 3 and Supplementary Fig. 7) and exhibit field-independent dip cascades, distinct from the QOs (see Supplementary Fig. 7**b**) and the Yamaji effect[40]. These field-independent AMROs are closely related to the commensurability resonance, also known as the Lebed magic-angle effect, which occurs in quasi-1D organic salts with open Fermi sheets[41-43]. We also note that AMRO signatures in ZrTe$_5$ have been reported in the literature[44,45], but the specific sequence of AMRO dips is reported here for the first time. The physical origin remains unclear up to now. The main feature of commensurability resonance is that the AMROs develop sharp dips at magic angles, which are periodic in tangent scale formulated as $\tan\theta = \frac{p}{q} * \frac{c}{b}$ with $p$, $q$ are integers and c, b are crystal constants of ZrTe$_5$. The commensurability resonance can be intuitively viewed as a geometric effect arising from the interaction between electron orbits oriented by the magnetic field and the underlying lattice[42,46]. Resonance occurs when the orbits commensurately ($p/q$) align with the lattice unit cell, forcing a series of resistance dips due to the condition that mini-suffering from velocity averaging is satisfied at the magic angle. However, the physics of the commensurability resonance in quasi-1D systems is found to be both deeper and concise, as even two layers are sufficient to induce it via tunneling-induced quantum interference, according to recent progress[47,48]. A transfer integral ($t_{pq}$) governs this geometric resonance, if **b** is the less conductive direction (the smallest hopping occurs along the out-of-plane **b** direction in ZrTe$_5$), the main non-zero elements are $q = 1$, then $p/q$ is always an integer $n$, fractions are rare except for special cases like magnetic-field-induced density wave[49]. As shown in the inset of Fig. 4**a**, the angles of large dips are rescaled to $\tan\theta$, and p/q are assigned to integers. Linear fitting produces a value of c/b ~ 0.951, which is close to the experimental value c/b = 0.945[21]. Unexpectedly, subunity fractions are observed in the commensurability resonance (marked by red vertical lines in Fig. 4**a**), which is incompatible with the anisotropy analysis in ZrTe$_5$ as mentioned above (fractional AMRO dips can be reproduced across different samples; see Supplementary Note 4 and Supplementary Fig. 8). Figure 4**c** shows the temperature-dependent AMRO dips: the integer fractions smear above 100 K, and the subunity fractions disappear more quickly. The amplitudes of integer and subunity fractional AMRO dips are extracted and summarized in Fig. 4**d** (see Supplementary Note 2 and Supplementary Fig. 4 for details of background subtraction). We can see that the temperature-dependent amplitudes of integer dips show slower damping than those of subunity fractions, which almost disappear around 25 K, which coincides with the onset temperature of mini-oscillations, $T_{\text{onset}}^{\text{mini}} \sim 25$ K (Fig. 2**b**), indicating the mini-oscillations correlate with the commensurability resonance. In Fig. 4**e**, all subunity fractions are summarized, and the main cascades exhibit 1/5 multiples, except for two fractions with 1/10 multiples.

However, there is one apparent inconsistency: commensurability resonance in quasi-1D materials requires an open Fermi surface (corrugated Fermi sheets), which is absent in ZrTe$_5$ reported in the literature[24]. Because Fermi sheets in quasi-1D materials do not contribute to SdH oscillations, the AMRO shown here might provide key evidence for their existence. However, the case here is quite different: the material is under an intensive magnetic field (in the quantum limit), relying solely on LBs, which invalidates the usual Fermi-surface-based (weak-field limit) analyses. The 1/5 multiple fractions indicate there are superstructure modulations along the **b** direction (b' = 5b), and then the Brillouin zone folding gives rise to mini-Landau bands gapped along the **b** direction. The gap along the **b** direction forces the cyclotron orbitals to lie strictly in the **ac** plane, satisfying the quantum-interference condition for AMROs and possibly explaining the observed commensurability resonance in ZrTe$_5$. Furthermore, the presence of mini-Landau bands along the **b** direction is directly related to the observed mini-oscillations, but how these mini-Landau bands respond to the magnetic field and generate the four sequential QO frequencies remains unknown. Another important observation is that the multiple subunity fractions are observed only in the quantum limit (> 1 T), indicating that the superstructure modulations are magnetic-field-induced, pointing to a possible magnetic-field-induced density wave or a novel magnetic field-induced structural transition[50]. Therefore, our observation warrants further exploration of field-induced structural modulations in ZrTe$_5$[50], which might be visualized by spectroscopic experiments under a magnetic field.

In conclusion, we detected sequential mini-oscillations in the quantum limit of ZrTe$_5$. These mini-oscillations represent an unknown pattern of quantum oscillations, related to the main oscillations and observed commensurability resonances. Furthermore, we established the Dirac semiconductor ZrTe$_5$ as a simple platform for exploring exotic QOs, alongside mainstream research on correlated insulators.

## Methods

### Crystal growth

Single crystals of ZrTe$_5$ were synthesized using Te-flux and chemical vapor transport (CVT) methods[14]. For the flux-grown samples, Zr (Alfa, 99.95%) and Te (99.9999%) were mixed in a 1:300 molar ratio, sealed in a quartz tube, and heated in a box furnace. The mixture was brought to 900 °C, periodically shaken, and maintained at that temperature for two days. Subsequently, the temperature was lowered to 660 °C over 7 hours, then slowly cooled to 460 °C over 200 hours. The ZrTe$_5$ crystals were separated from the Te flux by centrifuging at 460 °C. To grow larger crystals, iterative temperature cycling was applied. For CVT growth, ZrTe$_5$ powder was synthesized by reacting stoichiometric amounts of Zr (Aladdin, 99.5%) and Te, sealed with iodine in a quartz tube, and heated in a two-zone furnace at 540 °C and 445 °C for 1 month.

### Transport measurements

Electrical transport measurements up to 9 T were performed using a Quantum Design Physical Property Measurement System (PPMS–9 T) equipped with a home-built rotator. Measurements at fields up to 18 T at sub-Kelvin temperatures were conducted in a top-loading dilution refrigerator (Oxford TLM, base temperature approximately 20 mK). High-field measurements up to 35 T were performed in a water-cooled magnet at the Chinese High Magnetic Field Laboratory (CHMFL) in Hefei. Resistance was measured using SR830 lock-in amplifiers at frequencies of 13.333 Hz or 17.777 Hz. A Keithley 6221 AC/DC current source provided the current.

**Data availability** The data generated in this study are provided in the Source Data file. Additional data related to the current study are available from the corresponding author upon request. Source data are provided with this paper.

**Acknowledgments** We thank Yu Sun, Xinshan Tu, and Qizhe Chai for their assistance with transport measurements in an 18 T dilution fridge. C.-L. Zhang was supported by the National Key R&D Program of China (Grant No. 2024YFA1611102; Grant No. 2023YFA1407400), Beijing Natural Science Foundation (Z250005), and a start-up grant from the Institute of Physics, Chinese Academy of Sciences. J. Zhang was supported by the National Key R&D Program of the MOST of China (Grants No. 2022YFA1602602) and the National Natural Science Foundation of China (Grants No. 12474053). A portion of this work was carried out at the Synergetic Extreme Condition User Facility (SECUF).

**Author contributions** C.-L.Z. conceived and supervised the project. Z.W., D.X., S.P., B.T. and C.-L.Z. performed all the electrical transport experiments with the help from J.Z., G.L. and L.L.; D.X. and C.-L.Z. grew the single crystals; C.-L.Z., Z.W. and D.X. analyzed the data. C.-L.Z. and Z.W. wrote the paper with input from all other authors.

**Competing interests** The authors declare no competing interests.

**Figure Captions**

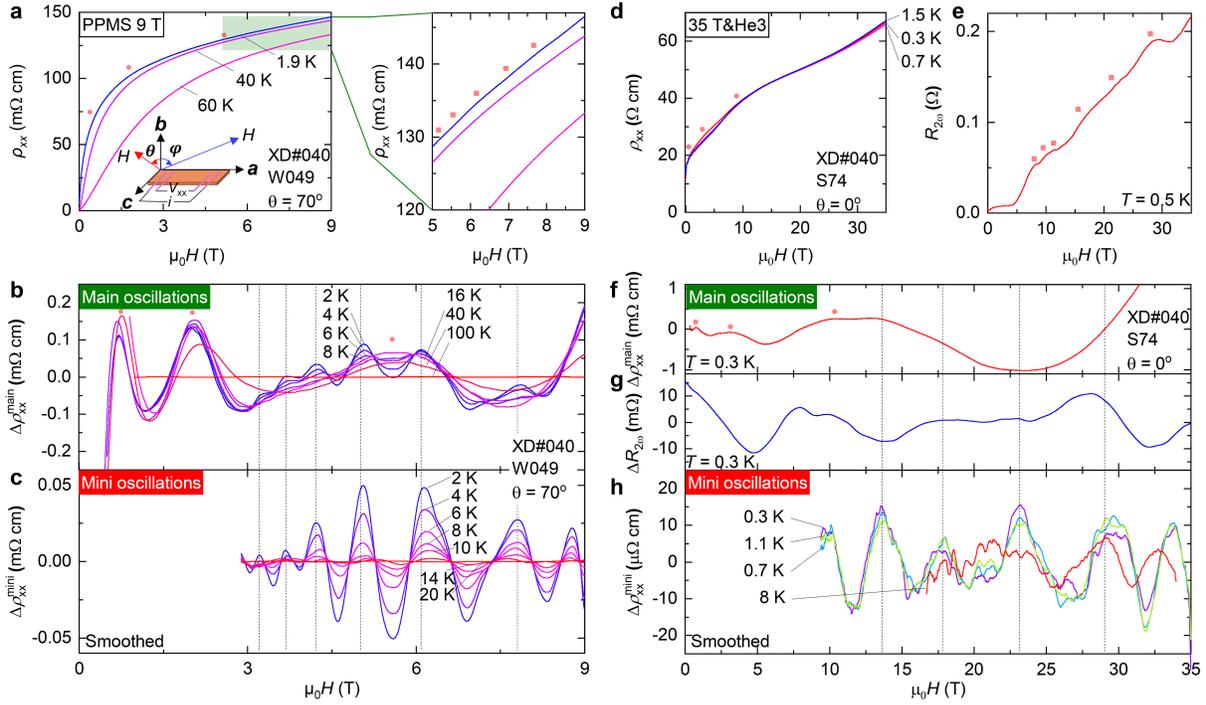

Fig. 1| **Observation of the mini-oscillations superimposed on the main quantum oscillations in the quantum limit. a,** Temperature-dependent magnetoresistance measured up to 9 T at θ = 70°, filled circles mark the peak positions of the main oscillations. The inset shows an illustration of the measurement setup, with the two defined tilted angles (θ & φ). The dashed green rectangle is the selected area enlarged to directly exhibit the mini-oscillations. **b,** Temperature-dependent main oscillations $\Delta\rho_{xx}^{main}$ with superimposed mini-oscillations. **c,** Temperature-dependent mini-oscillations $\Delta\rho_{xx}^{mini}$ obtained by further subtracting the slowly oscillating background due to the main oscillations (the curves are smoothed for extracting the amplitudes for L-K fittings). **d,** Temperature-dependent magnetoresistance is extended to 35 T at θ = 70°, filled circles indicate the peak positions of the main oscillations. The mini-oscillations are difficult to see with the eyes due to the limited signal resolution in the high-field facility. **e,** Second harmonic response $R_{2\omega}$ (equals to thermoelectric signals) clearly exhibits the mini-oscillations in the raw data measured up to 35 T. **f,** Background-subtracted $\Delta\rho_{xx}^{main}$ exhibits the main oscillations with mini-oscillations superimposed. **g,** Mini-oscillations on $\Delta R_{2\omega}$ with background subtracted.

**h,** Temperature-dependent $\Delta\rho_{xx}^{\text{mini}}$ obtained by further subtracting the slowly oscillating background from 10 to 35 T, and the subtracted data are also smoothed.

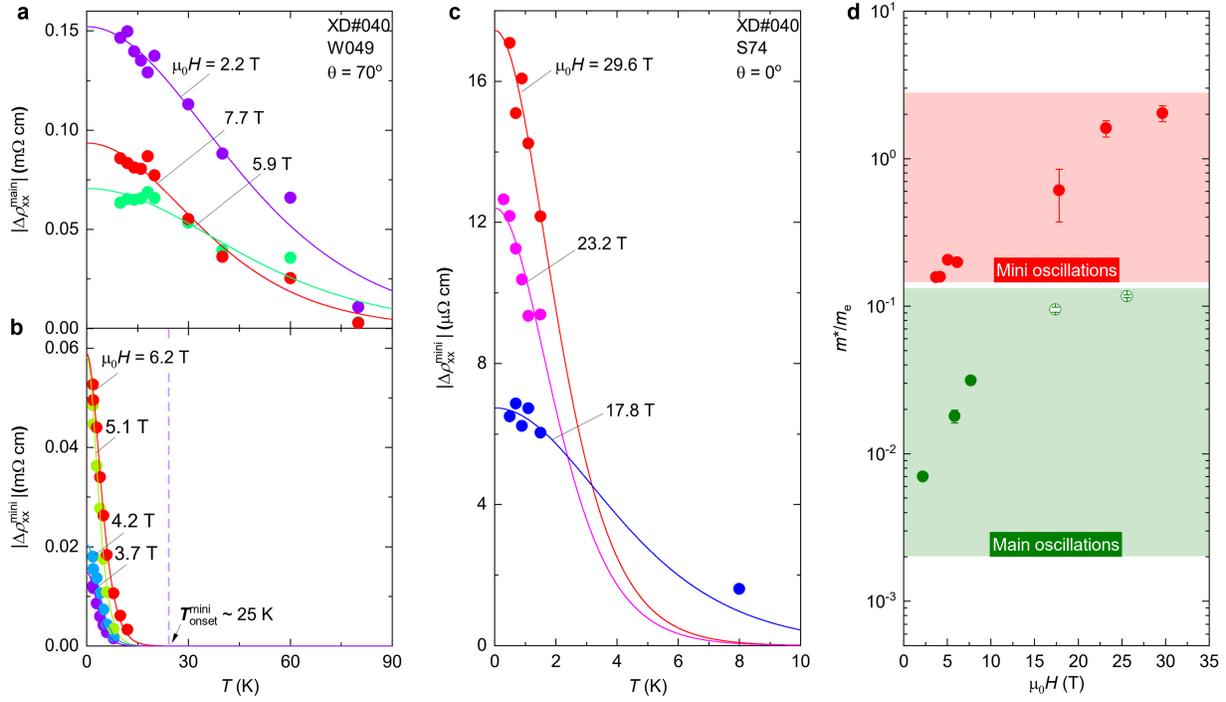

Fig. 2 | **Heavy effective mass of the mini-oscillations. a,** L–K formula fittings for the temperature-dependent amplitudes of $\Delta\rho_{xx}^{\text{main}}$ measured up to 9 T. **b,** L–K formula fittings for the temperature-dependent amplitudes of $\Delta\rho_{xx}^{\text{mini}}$ measured up to 9 T. Violet dashed vertical line shows the onset temperature of the mini-oscillations. **c,** L–K formula fittings for the temperature-dependent amplitudes of the mini-oscillations $\Delta\rho_{xx}^{\text{mini}}$ measured up to 35 T at He-3 temperatures. **d,** Summarized effective mass plotted against magnetic field, the effective mass of the mini-oscillations is one order heavier than that of the main oscillations. Data points below 10 T are from fittings $\Delta\rho_{xx}^{\text{mini}}$ at θ = 70°, while data points above 10 T are from fittings $\Delta\rho_{xx}^{\text{mini}}$ at θ = 0°. The two data points (unfilled circles) of the effective mass of the main oscillations are from Ref[14].

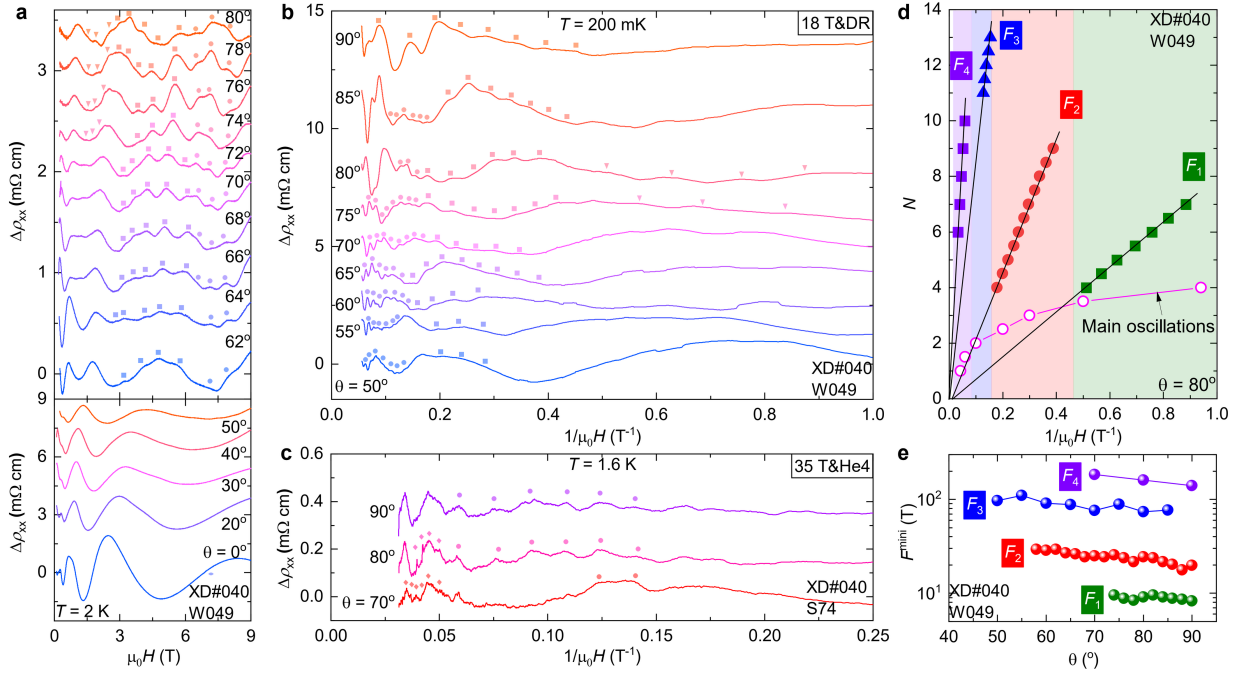

Fig. 3| **Sequential features of the mini-oscillations with high frequencies and its relation to the main oscillations in angle-dependent measurements. a,** Background-subtracted magnetoresistance $\Delta\rho_{xx}$ at different angles measured up to 9 T at 2 K. When the magnetic field is tilted towards the *c* axis, the mini-oscillations appear like peak splittings around 62°. Three sets of mini-oscillations marked by inverted triangles, rectangles and circles, respectively. **b,** Background-subtracted magnetoresistance $\Delta\rho_{xx}$ at different angles measured up to 18 T at 200 mK (raw data are smoothed). A full set of sequential mini-oscillations is uncovered, denoted by different markers. The mini-oscillations (several angles) with the range 1-3 T are not readily detected due to the spikes noise at low fields in the dilution refrigerator, but the oscillations correspond to $F_1$ can be detected in PPMS 9 T data (Fig. 3**a**), and further cross-checked as shown in Supplementary Fig. 9. **c,** Background-subtracted magnetoresistance $\Delta\rho_{xx}$ measured up to 35 T shows one additional frequency. **d,** Landau fan diagram for four sequential frequencies of the mini-oscillations and the set of main oscillations at the same angle, with arbitrary integers assigned to the peaks, respectively. Four-colored regimes indicate the sequential features and lock to each period of the main oscillations. **e,** Angle dependence of the four frequencies of the mini-oscillations.

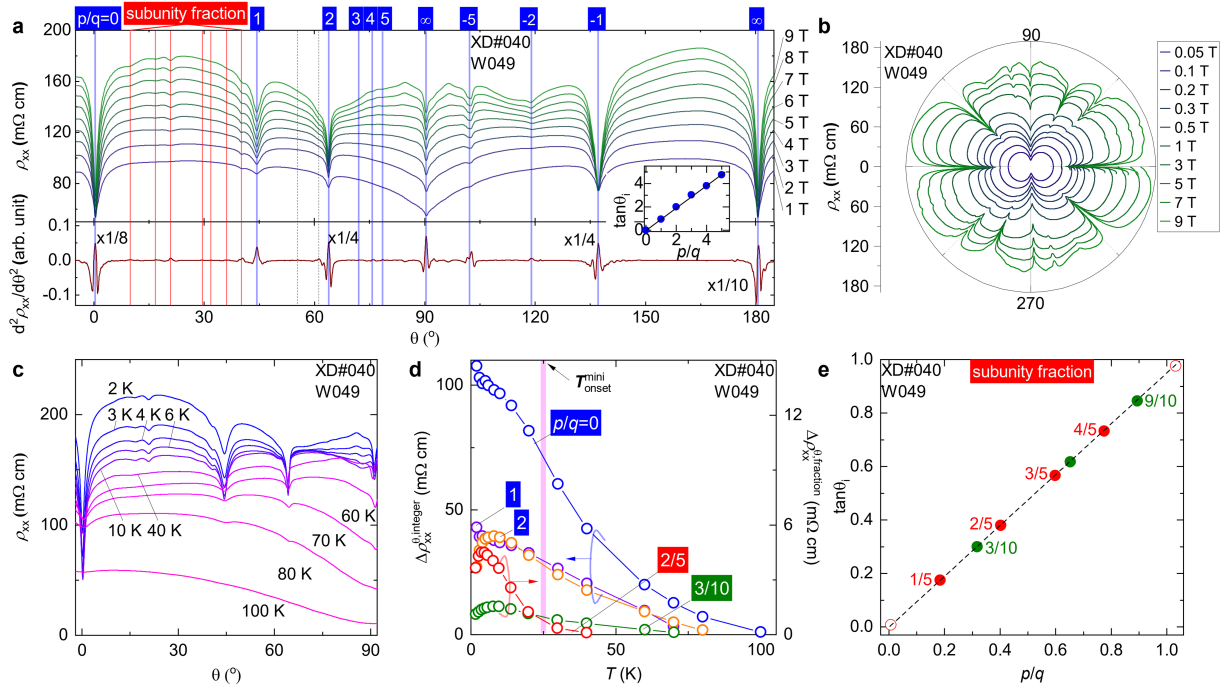

Fig. 4| **Subunity fractional magic angles in the commensurability resonance and relation to the observed mini-oscillations. a,** Angular magnetoresistance oscillations (AMROs) appear as field-independent sharp dips in resistance observed across different magnetic fields at 2 K. The inset shows the rescaled dip positions ($\tan\theta_i$) versus integer ($p/q$), fitted by the Lebed magic-angle relation $\tan\theta_i = \frac{p}{q} * \frac{c}{b}$ (zero-shift of $\theta$ is corrected). The lower panel shows the second derivative of the AMROs. **b,** AMROs measured at different fields in a polar plot. **c,** Temperature-dependent AMROs measured up to 100 K. **d,** Renormalized temperature-dependent amplitudes of AMROs at integers and fractions. The vertical eye-guided line shows the onset temperature of mini-oscillations. **e,** Subunity fractions mainly exhibit 1/5 cascades, with two exceptions that exhibit 1/10 multiples.

# Supplementary Information

# Observation of sequential quantum oscillations induced by mini-Landau bands in a three-dimensional Dirac semiconductor

Zezhi Wang[1,2*], Dong Xing[1,2*], Bingbing Tong[1], Senyang Pan[3], Guangtong Liu[1,4], Li Lu[1,4], Jinglei Zhang[3], and Cheng-Long Zhang[1†]

[1]*Beijing National Laboratory for Condensed Matter Physics, Institute of Physics, Chinese Academy of Sciences, Beijing 100190, China*

[2]*School of Physical Sciences, University of Chinese Academy of Sciences, Beijing 100049, China*

[3] *High Magnetic Field Laboratory, HFIPS, Chinese Academy of Sciences, Hefei 230031, China*

[4]*Hefei National Laboratory, Hefei 230088, China*

Corresponding author: chenglong.zhang@iphy.ac.cn

## Supplementary Note 1. Processes of background subtraction for obtaining the main and mini-oscillations $\Delta\rho_{xx}$

As shown in Supplementary Fig. 3**a**, the magnetoresistance increases rapidly at low fields. Therefore, we fitted the raw data by using the exponential functions $c * e^{-(x-x_0)/t_1}$. The main oscillations $\Delta\rho_{xx}^{main}$ were obtained by subtracting the fitted background (Supplementary Fig. 3**b**).

To clearly resolve the mini-oscillations, the raw data were further smoothed (Supplementary Fig. 3**c** and 3**d**). We fitted the slow background (due to the main oscillation) over the range 2.5 T to 9 T by using an 8th-order polynomial. The mini-oscillations $\Delta\rho_{xx}^{mini}$ were obtained by subtracting the fitted background (Supplementary Fig. 3**e** and 3**f**).

## Supplementary Note 2. Procedures for subtracting the background of AMROs

Temperature-dependent amplitudes of angular magnetoresistance oscillations (AMROs) dips are not straightforward to be quantitatively analyzed due to the strongly temperature-dependent background. To show the exact temperature-dependent amplitudes of AMRO dips, we subtracted the backgrounds for integer and fractional AMRO dips, respectively. As shown in Supplementary Fig. 4**a**, a point $(x_1, y_1)$ is selected around a relatively flat regime, and $y_1$ is used as the background of the dip. The lowest point of each dip $(x_0, y_0)$ is chosen, and the amplitudes of the integer AMROs dips are $|y_1 - y_0|$. As shown in Fig. 4**b**, due to no relatively flat regions near the dips, we selected two local maxima near each dips, $A_1(B_1)$ and $A_2(B_2)$, and the amplitudes of the fractional AMRO dips is then obtained by the vertical distance from the minimum point $A(B)$ of the dip to the line connecting $A_1(B_1)$ and $A_2(B_2)$.

**Supplementary Note 3. AMROs of CVT samples measured in *ba* plane**

As shown in Supplementary Fig. 7**b**, we measured AMROs in the *bc* plane of a CVT sample. AMROs of the CVT sample exhibit field-dependent AMROs that differ from those of flux-grown samples. According to the Lifshitz–Kosevich (L-K) formula of the Shubnikov-de Haas (SdH) oscillations[1]:

$$\frac{\Delta\rho(B)}{\rho_0} \propto R_T R_D R_s \cos\left(2\pi\left(\frac{B_f}{B}+\gamma\right)\right),$$

where $R_T$, $R_D$, $R_s$ are the temperature, Dingle, and spin damping factors, respectively, $B_f$ is the frequency and $\gamma$ is the phase of the SdH oscillations. The frequency of the SdH oscillation $B_f$ is determined by the extremal cross-sectional area $S_f$ of the Fermi surface that is perpendicular to the field direction. For an isotropic spherical Fermi surface, the extremal cross-sectional area is independent of field orientations. As a result, $B_f$ does not change with the rotation of field, and therefore no AMRO appears.

For a 2D cylindrical Fermi surface parallel to the *b* axis, $B_f^{2D} = B_{f,b}/cos\theta$, is the frequency when the magnetic field is along the *b* axis. The L–K formula can be simplified to:

$$\frac{\Delta\rho(B)}{\rho_0} \propto R_T R_D R_s \cos\left(2\pi\left(\frac{B_{f,b}}{Bcos\theta}+\gamma\right)\right),$$

where $\theta$ is the angle between the *B* and *b* axis. The maximum in $\rho_{xx}$ occurs when $B_{f,b}/Bcos\theta + \gamma = n$ with integer $n$. Hence, at the specific angle $\theta^*$, the Fermi level crossing the Landau bands causes AMROs. Supplementary Figure 7**c** shows the peak positions shift with the magnetic field given by

$$cos\theta^* = \frac{B_{f,b}}{B(n-\gamma)}.$$

However, the Fermi surface of metallic ZrTe$_5$ samples is a 3D ellipsoidal object[2], and SdH oscillations are different from those of a cylindrical Fermi surface. For a 3D ellipsoidal Fermi surface, when the magnetic field rotates in the **b-c** plane, $B_f^{3D}$ is given by,

$$B_f^{3D} = \frac{B_{f,b} B_{f,c}}{\sqrt{(B_{f,b}\sin\theta)^2 + (B_{f,c}\cos\theta)^2}}.$$

The fitting is approximately linear only at small angles in a system with large anisotropy, explaining the deviation from linearity at higher tilted angles, as shown in Supplementary Fig. 7**c**.

**Supplementary Note 4. AMROs of $\rho_{xx}$ for other flux-grown samples.**

To reproduce the AMRO dips, including the integer and fractional dips in flux-grown samples, we performed AMR measurements in the **bc** plane for different samples and batches. As shown in Supplementary Fig. 8**a**, all integer-dips occurred in the same position for all the measured samples. However, compared to the W049 sample, the integer AMRO dips in other samples are much smaller, and the fractional AMRO dips are not apparent in the raw data. To analyze the fractional AMRO dips, we performed background subtraction of the raw data using a large-window smooth as shown in Supplementary Fig. 8**b**. Supplementary Figure 8**c** shows $\Delta\rho_{xx}$ obtained by subtracting the smoothed background from the raw data. The fractional AMRO dips can be identified through $\Delta\rho_{xx}$, corresponding to those in W049. However, in other samples, the fractional AMRO dips are difficult to identify due to weaker amplitudes or background variations in this region.

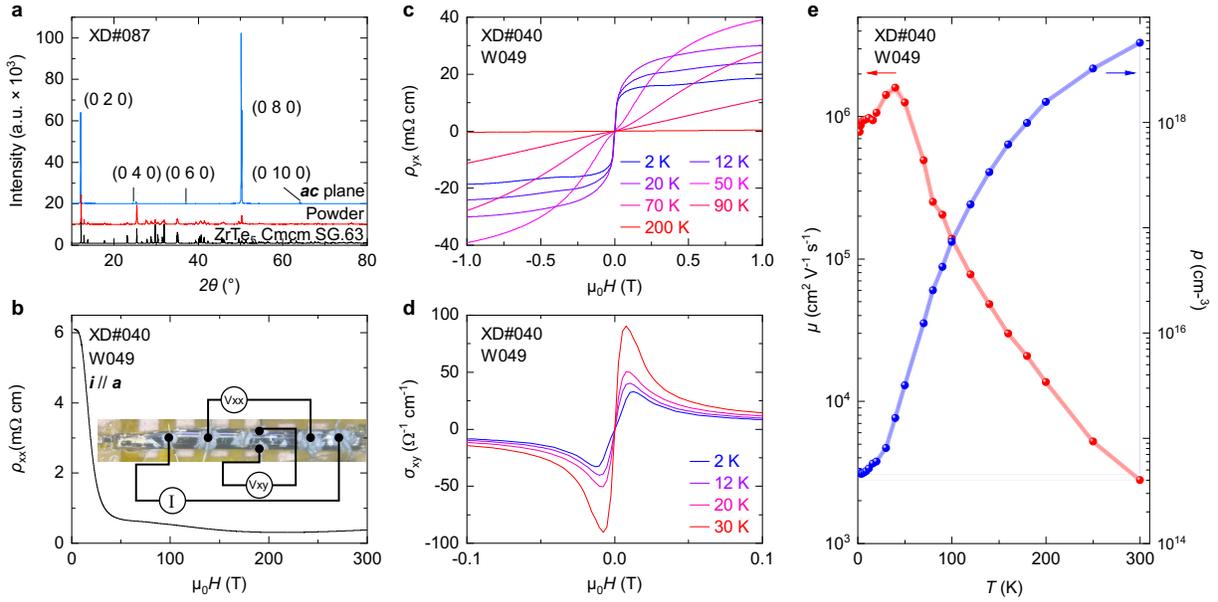

**Supplementary Fig. 1: Structural and electrical transport characterizations of ZrTe$_5$.** **a**, X-ray diffraction characterizations on powder and exfoliated samples, respectively. **b**, Temperature-dependent resistivity $\rho_{xx}$ of flux-grown ZrTe$_5$ sample. The inset is a photo of a typical device. **c**, Temperature-dependent Hall resistivity $\rho_{yx}$ of sample W049. **d**, Temperature-dependent Hall conductivity $\sigma_{xy}$. **e**, Temperature-dependent carrier density ($p$) and mobility ($\mu$) of the sample W049.

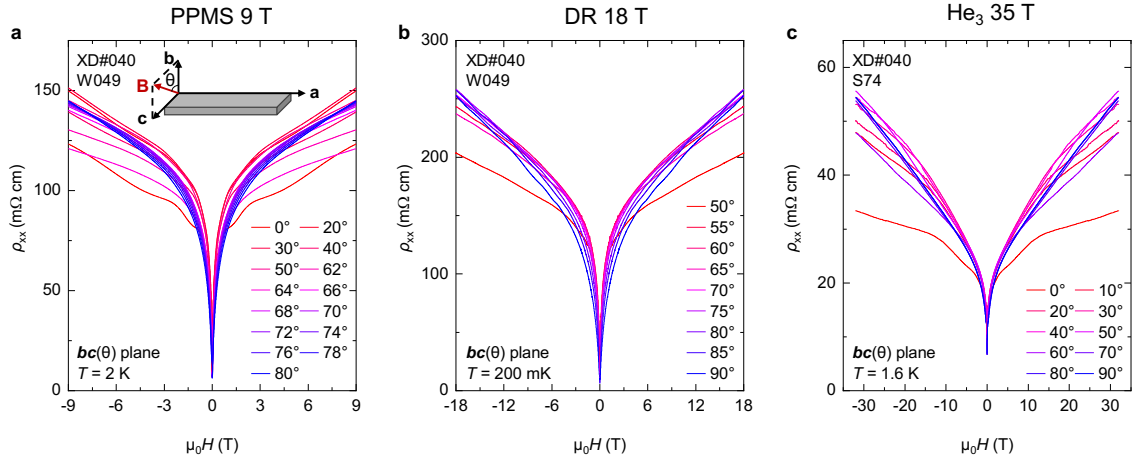

**Supplementary Fig. 2: Raw data of Mangtoresistance at different angles in *bc* plane.**
**a**, Magnetoresistance $\rho_{xx}$ at different angles in the *bc* plane of sample W049 measured up to 9 T at 2 K. The inset illustrates the experimental setup. $\theta = 90°$ corresponds to the magnetic field along *c* axis. **b**, Magnetoresistance $\rho_{xx}$ at different angles in the *bc* plane of sample W049 measured up to 18 T at 200 mK. **c**, Magnetoresistance $\rho_{xx}$ at different angles in the *bc* plane of sample S74 measured up to 35 T at 1.6 K.

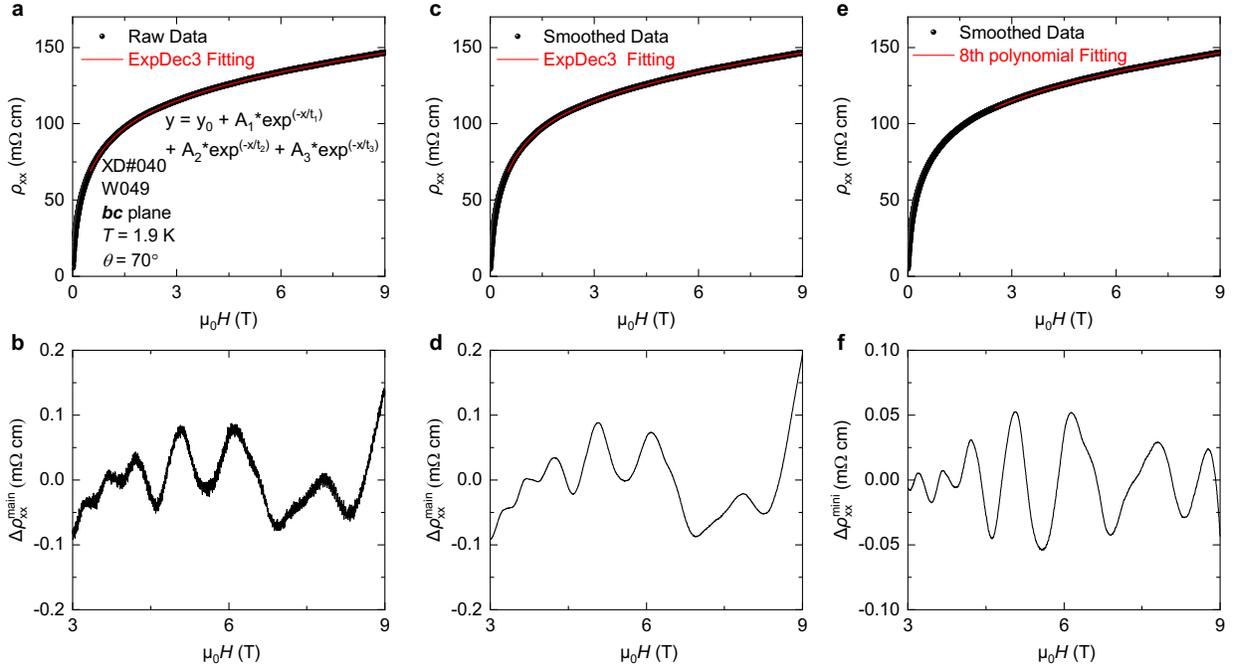

**Supplementary Fig. 3: Background subtraction for main and mini-oscillations $\Delta\rho_{xx}$.**
**a**, Raw data of W049 measured at 1.9 K with the magnetic field oriented at 70° within the *bc* plane. The red line is the fitted curve obtained using the ExpDec3 function. **c**, Data smoothed from the raw data. The red line is the fitted curve obtained using the ExpDec3 function. **e**, The red line is the fitted curve obtained using the 8th-order polynomial. **b-f**, The oscillatory component $\Delta\rho_{xx}$ obtained by subtracting the fitted data from the raw data in Supplementary Figs. 3**a-e**, respectively.

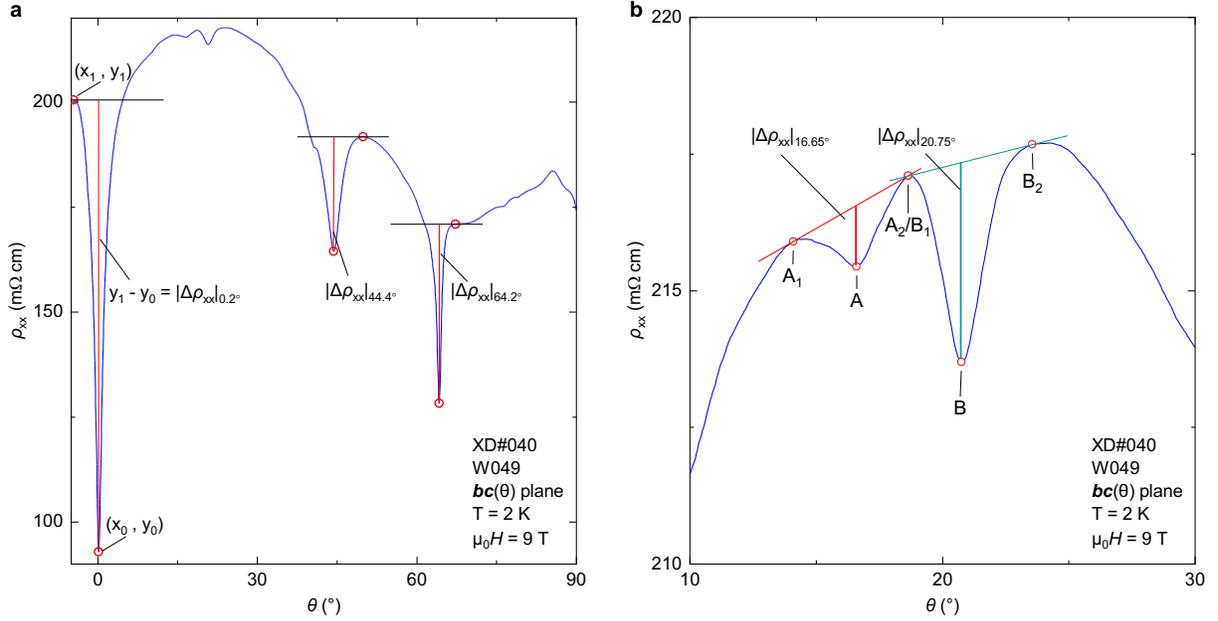

**Supplementary Fig. 4: The characterization of the amplitudes of integer and fractional AMRO dips.** Blue line is the AMR raw data of the W049 sample within the ***bc*** plane at 9 T. **a**, The characterization of the amplitudes of three integer AMRO dips. For each dip, select the ($x_0$, $y_0$) as the lowest point where $y_0$ is considered the peak value of the dip. A point ($x_1$, $y_1$) is chosen from the relatively flat region near the dip as the background. The difference $|\Delta\rho_{xx}|_\theta = |y_0 - y_1|$ is defined as the amplitude of the integer AMRO dip. **b**, The characterization of the amplitude of two fractional AMRO dips. For each dip, select the middle point of the dip as the lowest point of $A(B)$. $A_1(B_1)$ and $A_2(B_2)$ are two points selected from either side of the dip, and the straight line connecting these two points is used as the background. The vertical distance from point $A(B)$ to this line is defined as the amplitude of the fractional AMRO dip.

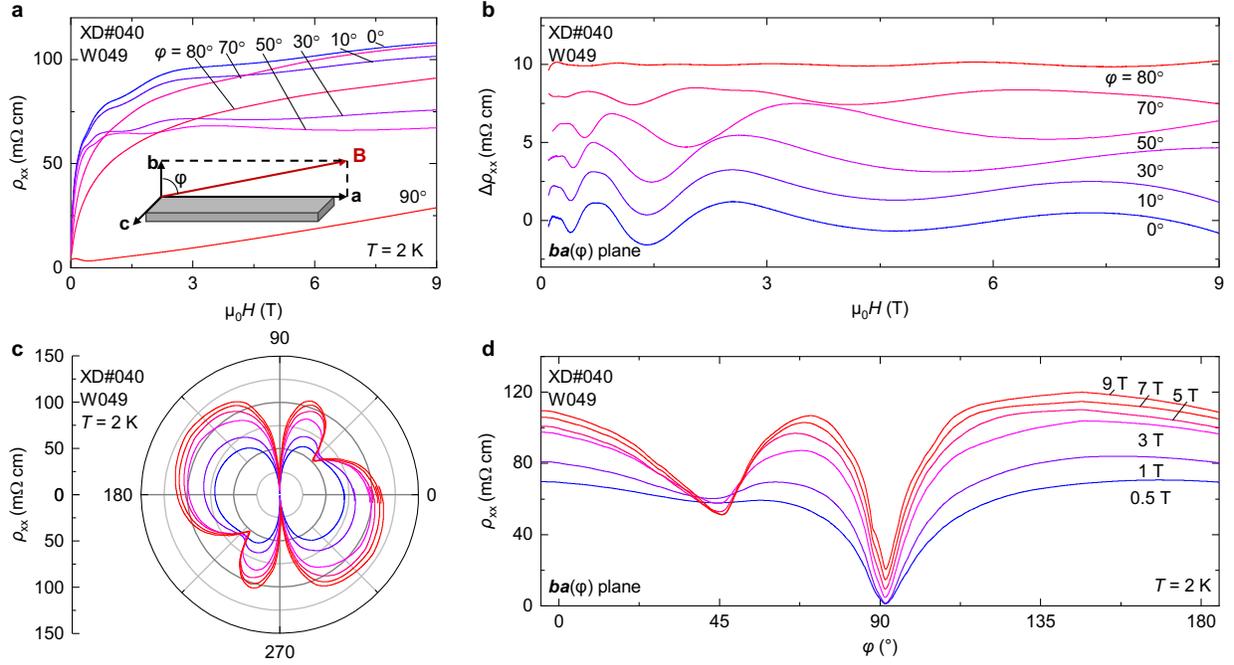

**Supplementary Fig. 5: Angle dependence of $\rho_{xx}$ in *ba* plane for sample W049. a**, $\rho_{xx}$ measured in the ***ba*** plane with rotating angle $\varphi$ (defined in the inset illustration). $\varphi = 90°$ corresponds to the magnetic field along ***a*** axis. **b**, Oscillatory component $\Delta\rho_{xx}$ of tilted angles $\varphi$ (***ba*** plane) measured at 2 K. **c & d**, The polar coordinate and linear coordinate plots of angle dependence of magnetoresistivity $\rho_{xx}$ measured in ***ba*** plane at different fixed fields at 2 K, respectively.

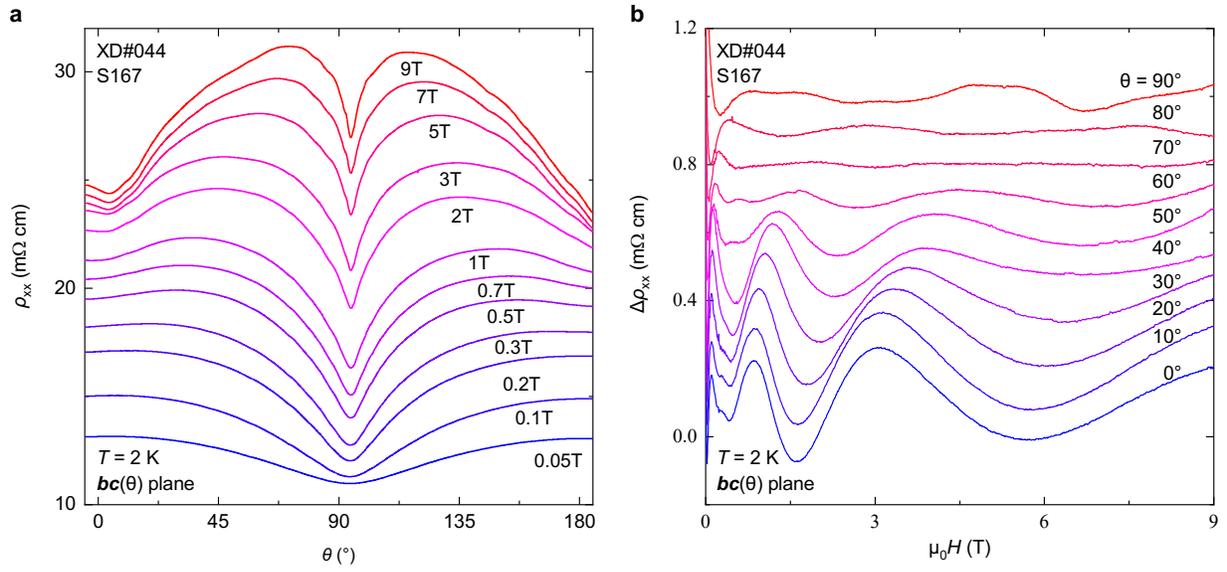

**Supplementary Fig. 6: Angle dependence of $\rho_{xx}$ in *bc* plane for sample S167. a**, Angle dependence of magnetoresistivity $\rho_{xx}$ measured in the *bc* plane at different fixed magnetic fields at 2 K. **b**, Oscillatory component $\Delta\rho_{xx}$ of tilted angles $\theta$ (*bc* plane) measured at 2 K.

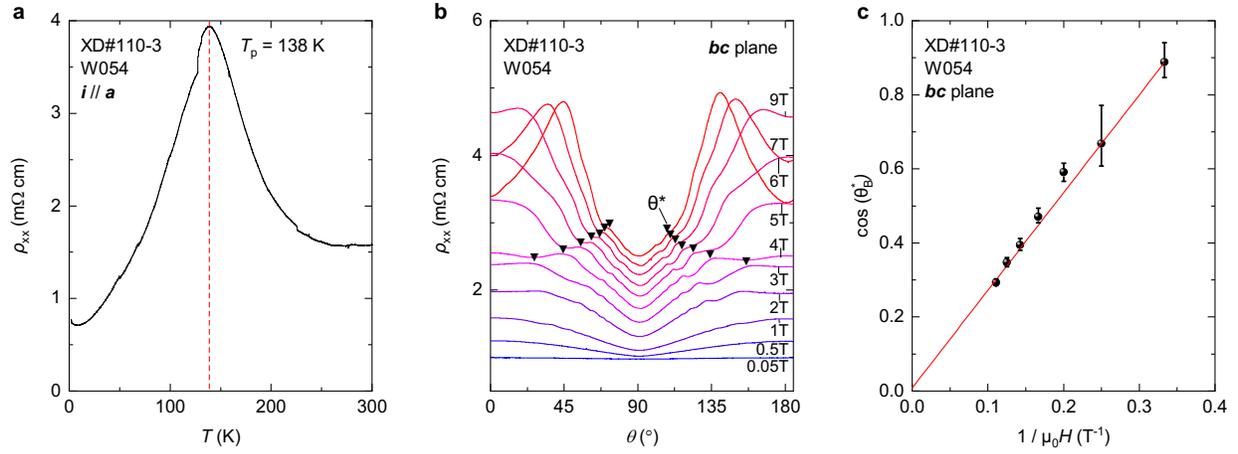

**Supplementary Fig. 7: Angle dependence of mangtoresistivity $\rho_{xx}$ in *bc* and *ba* planes for CVT samples. a**, Temperature-dependent resistivity $\rho_{xx}$ of the CVT sample W054. **b**, Angle-dependent magnetoresistivity $\rho_{xx}$ measured in the *bc* plane in constant fields at 2 K for W054. The inverted triangles mark the peak positions $\theta^*$. **c**, The dependence of $\theta^*$ on $B$ plotted as $\cos\theta^*$ versus $B^{-1}$.

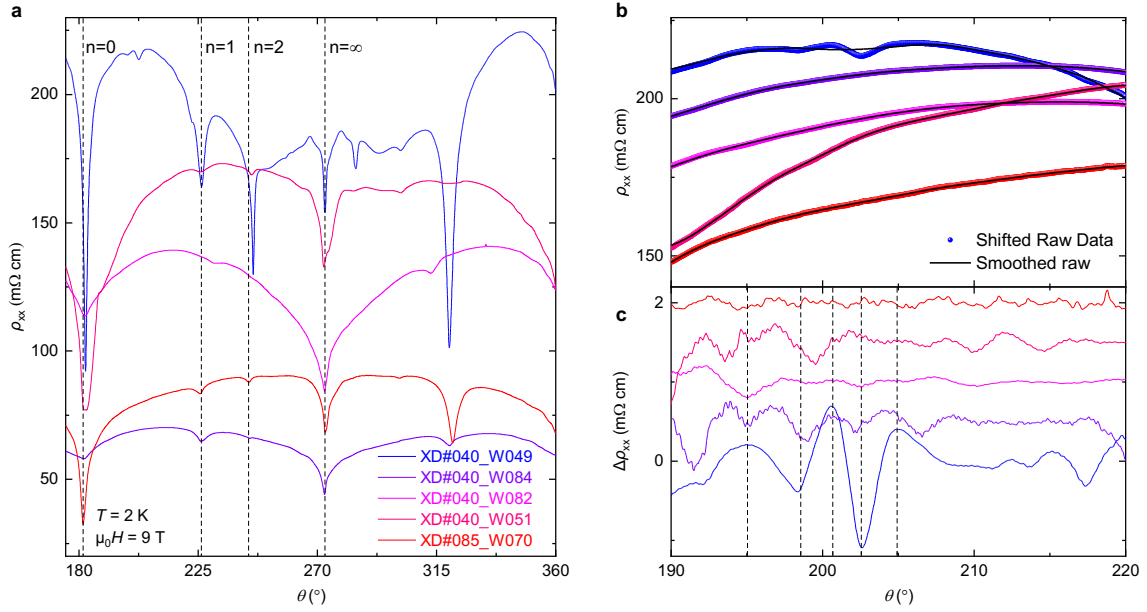

**Supplementary Fig. 8: Angle dependence of mangtoresistivity $\rho_{xx}$ for other flux-grown samples. a**, Angle dependence of mangtoresistivity $\rho_{xx}$ measured in the ***bc*** plane in 9 T at 2 K for different flux samples. The dashed lines indicate the positions of $\frac{p}{q} = n = b/c \cdot \tan(\theta) = 0, 1, 2, \infty$. **b**, Raw of AMR data measured in different samples, the black line represents the data smoothed using the Loess method. Curves are shifted for clarity. **c**, The oscillatory component $\Delta\rho_{xx}$ obtained by subtracting the smoothed data from the raw data in Supplementary Fig. 8**b**, respectively.

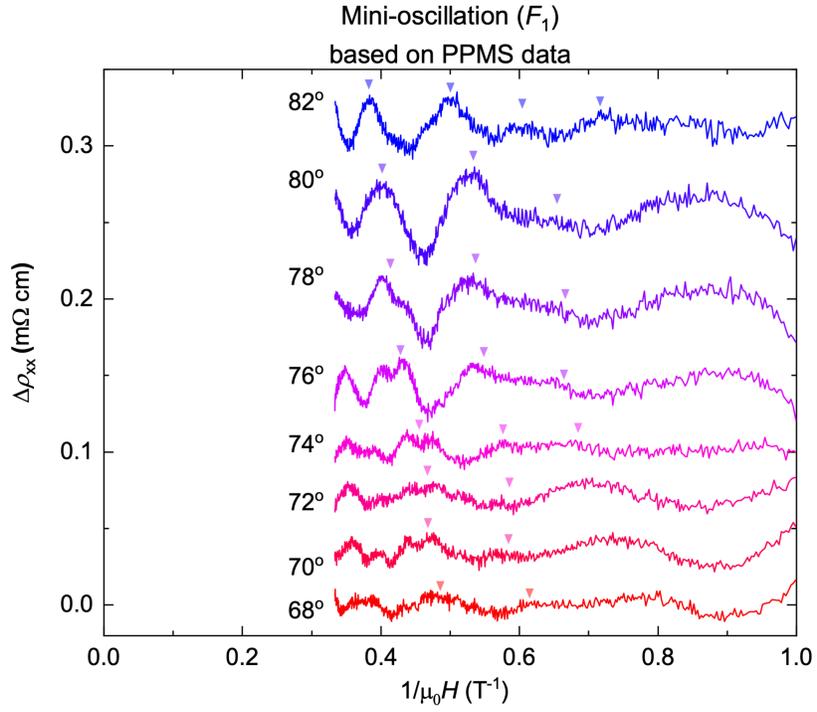

**Supplementary Fig. 9:** Cross-check analysis for frequency $F_1$ of the mini-oscillations based on PPMS data from Fig. 3a of the main text.

# Supplementary References